\documentclass[preprint,preprintnumbers,showpacs,aps,amssymb]{revtex4-1}

\usepackage{graphicx}
\usepackage{epsfig}
\usepackage{bm}
\usepackage{amsmath}

\def\MBH{M_{\rm BH}}
\def\tBH{\tau_{\rm BH}}

\def\td{{\tilde\delta}}
\def\tE{{\tilde E}}
\def\tL{{\tilde\Lambda}}
\def\tR{{\tilde R}}
\def\tw{{\tilde\omega}}

\def\wmax{\omega_{\rm max}}
\def\r0min{r_{0,{\rm min}}}

\def\calO{{\cal O}}

\def\nn{\nonumber}

\begin{document}
\title{Black-hole bombs at the LHC}
\author{Jong-Phil Lee}
\email{jplee@kias.re.kr}
\affiliation{Department of Physics and IPAP, Yonsei University, Seoul 120-749, Korea}
\affiliation{Division of Quantum Phases $\&$ Devices, School of Physics, Konkuk University, Seoul 143-701, Korea}

\begin{abstract}
A particle scattered off by a rotating black hole can be amplified when the system is in the superradiant regime.
If the system is surrounded by a mirror which reflects the particle back to the black hole the whole system forms a black-hole bomb, 
amplifying the original field exponentially.
We show in this paper that higher dimensional black holes can also form black-hole bombs at the LHC.
For a pion the $e$-folding time for the field amplification is $t_c\sim 10^{-23}-10^{-24}$ sec.
If the lifetime of the black hole is long enough compared with $t_c$, we can observe severely amplified fields.
\end{abstract}

\maketitle
\section{Introduction}
Successful runnings of the Large Hadron Collider (LHC) at CERN are now opening a new era of high energy physics.
It is highly anticipated that not only the last missing piece of the Standard Model (SM) of particle physics, the Higgs boson, would be
discovered, but the new physics beyond the SM would appear for the first time.
One of the most striking events that the LHC would reveal is the production of mini black holes.
Mini black holes can be produced when there are extra dimensions \cite{XD}.
In these scenarios the fundamental energy scale for strong gravitation $M_D$ in $D=(4+n)$-spacetime dimensions is much smaller
than the Planck mass.
In the case of $M_D\sim 1$ TeV, the LHC experiments can probe the strong gravity regime and possibly produce black holes \cite{BH}.
Once the black hole is produced it decays instantaneously through Hawking radiation \cite{Hawking}.
Typically the lifetime of a mini black hole is $\sim\calO(10^{-26})$ sec and the size of the Schwarzschild radius is 
$\sim\calO(10^{-4})$ fm.
If the black hole is produced at the LHC, we can detect its creation by observing Hawking radiation.
In these reasons there have been many works on the emission of particles by mini black holes both static and rotating 
in extra dimensions \cite{Frolov,Ida,Harris,Pappas}. 
\par
Rotating (or Kerr) black holes exhibit a very interesting phenomenon known as superradiant scattering.
When a wave of the form $e^{-i\omega t}e^{im\phi}$ is incident on a rotating object of angular velocity $\Omega$,
the wave is amplified if $\omega<m\Omega$.
If one surrounds the rotating object by a mirror which reflects the scattered wave, the wave bounces back and forth between
the black hole and the mirror.
During the process the wave amplifies itself and the extracted energy from the black hole grows exponentially.
This is so called the Press-Teukolsky black-hole bomb \cite{Press}.
Even though there were no mirror, a mass term can play a role of the mirror to reflect the scattered field back to the black hole.
In this case the massive field is in the bound state with the black hole, resulting in the instability of the black hole.
There are many works on the black hole bomb where a massive particle is bound to a celestial black hole under superradiance 
\cite{Cardoso,Dolan,Hod}.
In \cite{Yoshida}, however, it was shown that higher dimensional Kerr black holes do not exhibit such instabilities for massive fields
because there are no bound geodesics in the spacetime around the black hole.
\par
In this work we analyze the emission of a massive scalar field by a rotating mini black hole in higher dimensions.
To focus on the possibility of forming a black hole bomb, we simply impose the mirror condition as a boundary condition.
As shown in \cite{Yoshida}, the mass term is not enough for the role of mirror, 
and electromagnetic fields might do the work for charged particles,
but we do not consider any technical details of the mirror.
We simply assume that the mirror boundary condition is given by whatever reflects the scattered particle back to the black hole.
The imaginary part of the field frequency $\omega$ which is responsible for the field amplification is then easily obtained by matching the
solutions of the Klein-Gordon equation and by the boundary condition.
It is possible that particles are emitted from the higher dimensional black hole not only on the brane but also in the bulk.
In general, the relative ratio of bulk-to-brane energy emissivity is quite less than unity \cite{Kanti}, as is expected intuitively.
More specifically, the imaginary part of $\omega$ for the bulk emission is suppressed by a factor of 
$\sim(\tw_*)^n$  compared with the brane emission where $\tw_*$ is a small quantity, so we do not consider the bulk emission
in the present work.
\section{Scalar emission from the rotating black holes in extra dimensions}
In this section, we describe the rotating black holes in higher $4+n$ dimensions  just as in the way of \cite{Pappas}.
We assume that the angular momentum lies parallel to our brane, and the black hole emits particles only on the brane.
A proper background geometry is \cite{Pappas}
\begin{eqnarray}
 ds^2
&=&
-\left(1-\frac{\mu}{\Sigma~r^{n-1}}\right)dt^2 -\frac{2a\mu\sin^2\theta}{\Sigma~r^{n-1}}dtd\phi
+\frac{\Sigma}{\Delta}dr^2 \nn\\
&&
+\Sigma d\theta^2+\left(r^2+a^2+\frac{a^2\mu\sin^2\theta}{\Sigma~r^{n-1}}\right)\sin^2\theta d\phi^2~,
\label{metric}
\end{eqnarray}
where
\begin{equation}
 \Delta=r^2+a^2-\frac{\mu}{r^{n-1}}~,~~~\Sigma=r^2+a^2\cos^2\theta~.
\end{equation}
Here $\mu$ is related to the black hole mass $\MBH$ and the fundamental mass scale $M_D$ in $D=4+n$ dimensions via
\begin{equation}
 \mu=\frac{\Gamma(\frac{n+3}{2})}{(n+2)\pi^{\frac{n+3}{2}}} (2\pi)^n \frac{\MBH}{M_D^{n+2}}~,
\end{equation}
and $a$ is proportional to the black hole angular momentum $J$ as
\begin{equation}
 J=\frac{2}{n+2}\MBH a~.
\label{J}
\end{equation}
The black hole horizon is determined by the condition $\Delta=0$, and the horizon radius $r_h$ satisfies
\begin{equation}
 r_h^{n+1}=\frac{\mu}{(1+a_*^2)}~,
\end{equation}
where $a_*=a/r_h$.
\par
Now consider a massive scalar field of mass $m_0$ under the gravitational background (\ref{metric}).
The scalar field $\Phi$ satisfies the Klein-Gordon equation in curved spacetime
\begin{equation}
 \frac{1}{\sqrt{-G}}\partial_A\left(\sqrt{-G}G^{AB}\partial_B\Phi\right)-m_0^2\Phi=0~,
\end{equation}
where $G_{AB}$ is the $D$-dimensional metric tensor and $G$ is its determinant.
To solve the above equation, we try the separation of variables for $\Phi$ as
\begin{equation}
 \Phi=e^{-i\omega t}e^{im\phi}R(r)S(\theta)~.
\label{vs}
\end{equation}
The radial function $R(r)$ satisfies 
\begin{equation}
 \frac{d}{dr} \left(\Delta\frac{dR}{dr}\right)+\left(
 \frac{K^2}{\Delta}-\tL_{jm}-m_0^2r^2\right)R=0~,
\end{equation}
while the angular function $S(\theta)$ satisfies
\begin{equation}
 \frac{1}{\sin\theta}\frac{d}{d\theta}\left(\sin\theta\frac{dS}{d\theta}\right) +\left(
 \tw^2a^2\cos^2\theta-\frac{m^2}{\sin^2\theta}+\tE_{jm}\right)S=0~,
\end{equation}
where
\begin{eqnarray}
 \tw&=&\sqrt{\omega^2-m_\phi^2}~,\\
 K&=&(r^2+a^2)\omega-am~,\\
 \tL_{jm}&=&\tE_{jm}+a^2\omega^2-2am\omega~,
\end{eqnarray}
and $\tE_{jm}$ is expanded around small $a\tw$ as
\begin{equation}
\tE_{j\ell m}=j(j+1)+(a\tw)^2\frac{-2j(j+1)+2m^2+1}{(2j-1)(2j+3)}+\calO((a\tw)^4)~.
\end{equation}
\par
Since the absorption and reflection probabilities of scattering are determined by the radial function, 
we concentrate on $R(r)$.
In the near-horizon region ($r\simeq r_h$), we change the variable as
$r\to f(r)=\Delta(r)/(r^2+a^2)$, and the radial equation becomes 
\begin{equation}
 f(1-f)\frac{d^2R}{df^2}+(1-D_*f)\frac{dR}{df}+\left[
 \frac{K_*^2}{A_*^2f(1-f)}-\frac{C_*}{A_*^2(1-f)}\right]R=0~,
\end{equation}
where
\begin{eqnarray}
 A_*&=&(n+1)+(n-1)a_*^2~,\\
K_*&=&(1+a_*^2)\omega r_h-a_*m~,\\
C_*&=&(\tL_{jm}+m_0^2r_h^2)(1+a_*^2)~,\\
D_*&=&1+\frac{n(1+a_*^2)}{A_*}-\frac{4a_*^2}{A_*^2}~.
\end{eqnarray}
The solution for the near-horizon region is proportional to the hypergeometric function $F(a,b,c;f)$ as
\begin{equation}
 R_{\rm NH}(f)=A_-f^\alpha(1-f)^\beta F(a,b,c;f)~,
\label{RNH}
\end{equation}
where 
\begin{eqnarray}
 \alpha&=&-i\frac{K_*}{A_*}~,\\
 \beta&=&\frac{1}{2}\left[(2-D_*)-\sqrt{(2-D_*)^2-\frac{4(K_*^2-C_*)}{A_*^2}}~\right]~,
\end{eqnarray}
with
\begin{equation}
 a=\alpha+\beta+D_*-1~,~~~b=\alpha+\beta~,~~~c=1+2\alpha~,
\end{equation}
and $A_-$ is the integration constant.
\par
Nextly, in the far-field region ($r\gg r_h$), the radial equation becomes
\begin{equation}
 \frac{d^2\tR}{dz^2}+\frac{1}{z}\frac{d\tR}{dz}+\left[
 1-\frac{\tE_{jm}+a^2\tw^2+1/4}{z^2}\right]\tR=0~,
\end{equation}
where $\tR(r)\equiv \sqrt{r} R(r)$ and $z\equiv \tw r$.
The solution of this equation is the Bessel function
\begin{equation}
 R_{\rm FF}(r)=\frac{B_1}{\sqrt{r}}J_\nu(\tw r)+\frac{B_2}{\sqrt{r}}Y_\nu(\tw r)~,
\label{RFF}
\end{equation}
where $\nu=\sqrt{\tE_{jm}+a^2\tw^2+1/4}$.
If we match the two solutions $R_{\rm NH}$ and $R_{\rm FF}$, we get the ratio of $B_2$ to $B_1$ as \cite{Pappas}
\begin{eqnarray}
 \frac{B_2}{B_1}
&=&-\pi\left[\frac{\tw r_h(1+a_*^2)^{1/(n+1)}}{2}\right]^{2j+1}
 \frac{1}{\nu\Gamma^2(\nu)}\nn\\
&&\times
 \frac{\Gamma(2\beta+D_*-2)\Gamma(2+\alpha-\beta-D_*)\Gamma(1+\alpha-\beta)}
        {\Gamma(\alpha+\beta+D_*-1)\Gamma(\alpha+\beta)\Gamma(2-2\beta-D_*)}~.
\label{match}
\end{eqnarray}
\par
Now we impose the boundary condition at the mirror location, $r=r_0$.
At the mirror, the scalar particle is bounced back to the black hole, thus we set $R_{\rm FF}(r_0)=0$, which means
\cite{Cardoso}
\begin{equation}
 \frac{B_1}{\sqrt{r_0}}J_\nu(\tw r_0)+\frac{B_2}{\sqrt{r_0}}Y_\nu(\tw r_0)=0~.
\label{BC}
\end{equation}
Combining Eqs.\ (\ref{match}) and (\ref{BC}) yields
\begin{eqnarray}
 \frac{J_\nu(\tw r_0)}{Y_\nu(\tw r_0)}
&=&\pi\left[\frac{\tw r_h(1+a_*^2)^{1/(n+1)}}{2}\right]^{2j+1}
 \frac{1}{\nu\Gamma^2(\nu)}\nn\\
&&\times
 \frac{\Gamma(2\beta+D_*-2)\Gamma(2+\alpha-\beta-D_*)\Gamma(1+\alpha-\beta)}
        {\Gamma(\alpha+\beta+D_*-1)\Gamma(\alpha+\beta)\Gamma(2-2\beta-D_*)}~.
\label{BC2}
\end{eqnarray}
In the limit of $\tw_*\equiv\tw r_h\ll 1$ the right-hand-side of Eq.\ (\ref{BC2}) is negligible, so
\begin{equation}
 \frac{J_\nu(\tw r_0)}{Y_\nu(\tw r_0)}\sim(\tw_*)^{2j+1}\simeq 0~.
\label{zero}
\end{equation}
(In the case of bulk emission, the ratio is proportional to $\sim (\tw_*)^{2j+1+n}$ \cite{Pappas},
as mentioned in the Introduction.)
One can conclude that the value of $\tw r_0$ is very close to the zeros of the Bessel function $J_\nu(x)$,
\begin{equation}
 \tw r_0=x_{\nu,k}~,
\end{equation}
where $J_\nu(x_{\nu,k})=0$.
We now assume that ${\rm Re}(\tw)\gg {\rm Im}(\tw)$, and the solution of Eq.\ (\ref{BC2}) can be written as 
\begin{equation}
 \tw\simeq \frac{x_{\nu,k}}{r_0}+\frac{i\td}{r_0}~,~~~|\td|\ll 1~,
\end{equation}
where we introduced $\td/r_0$ as a small imaginary part of $\tw$.
Since $J_\nu(\tw r_0)=J_\nu(x_{\nu,k}+i\td)\simeq i\td J'_\nu(x_{\nu,k})$, one arrives at the following expression:
\begin{eqnarray}
i\td
 &=&\frac{Y_\nu({x_{\nu,k}})}{J'_\nu(x_{\nu,k})}
\pi\left[\frac{\tw_*(1+a_*^2)^{1/(n+1)}}{2}\right]^{2j+1}
 \frac{1}{\nu\Gamma^2(\nu)}\nn\\
&&\times
 \frac{\Gamma(2\beta+D_*-2)\Gamma(2+\alpha-\beta-D_*)\Gamma(1+\alpha-\beta)}
        {\Gamma(\alpha+\beta+D_*-1)\Gamma(\alpha+\beta)\Gamma(2-2\beta-D_*)}~.
\label{td}
\end{eqnarray}
The imaginary part of the right-hand-side of the above equation comes from the terms containing $\alpha=-iK_*/A_*$.
The right-hand-side of Eq.\ (\ref{td}) also contains a sizable real part. 
It can be considered as a correction to $\tw r_0=x_{\nu,k}$, and we found that it is typically around 1\%, and neglected.
Using the properties of the Gamma functions, the imaginary part of Eq.\ (\ref{td}) is given by
\begin{eqnarray}
 \td
 &=&\frac{Y_\nu({x_{\nu,k}})}{J'_\nu(x_{\nu,k})}
\left[\frac{\tw_*(1+a_*^2)^{1/(n+1)}}{2}\right]^{2j+1}
\frac{1}{2}\sinh\pi\left(\frac{2K_*}{A_*}\right)\nn\\
 &&\times
\frac{\left[\beta^2+\frac{K_*^2}{A_*^2}\right]\left|\Gamma\left(-\beta+i\frac{K_*}{A_*}\right)\right|^2
\left[(1-\beta-D_*)^2+\frac{K_*^2}{A_*^2}\right]\left|\Gamma\left(1-\beta-D_*+i\frac{K_*}{A_*}\right)\right|^2}
{\nu\Gamma^2(\nu)(2-2\beta-D_*)\Gamma^2(2-2\beta-D_*)}\nn\\
&\simeq&
\frac{Y_\nu({x_{\nu,k}})}{J'_\nu(x_{\nu,k})}
\left[\frac{\tw_*(1+a_*^2)^{1/(n+1)}}{2}\right]^{2j+1}
\frac{\Gamma^2(1-\beta)\Gamma^2(2-\beta-D_*)}
{\nu\Gamma^2(\nu)(2-2\beta-D_*)\Gamma^2(2-2\beta-D_*)}
\left(\frac{\pi K_*}{A_*}\right)~,
\end{eqnarray}
where the last relation holds for small $K_*$.
One can easily check that the prefactor of $Y_\nu(x_{\nu,k})/J'_\nu(x_{\nu,k})$ is negative for our relevant range of $\omega$.
Thus for a small value of $K_*$, one has $i\td\sim -iK_*$.
The sign of $\td$ is opposite to that of $K_*=(1+a_*^2)\omega r_h-a_*m$.
It means that $\td>0$ for $(1+a_*^2)\omega r_h<a_*m$, which is the superradiant condition.
From Eq.\ (\ref{vs}), the time-dependent term is now proportional to 
\begin{equation}
 e^{-i\omega t}\sim e^{\delta t}~,
\label{amp}
\end{equation}
where $\delta=\td/r_0$.
Therefore, positive $\delta$ ensures the amplification of the outgoing field.
Note that the superradiance occurs only when $m>0$ and the minimum value of $j$ for the superradiance is $1$,
which guarantees the validity of our setting of (\ref{zero}).
Also, from Eq.\ (\ref{RFF}), the outgoing field at large $r$ is 
$R_{\rm FF}^{\rm out}\sim e^{i\tw r}/r\simeq e^{i(x_{\nu,k}/r_0)r} e^{-(\td/r_0)r}/r$, 
which vanishes (diverges) at $r\to\infty$ for positive (negative) $\td$.
Thus the system is in its bound state for $\td>0$ (or $\delta>0$).
\par
\section{Results and discussions}
We fix $M_D=1$ TeV, $\MBH=5$ TeV in this analysis.
The relevant range of $\omega$ for the superradiance is
\begin{equation}
 m_0<\omega<\frac{a_*m}{(1+a_*^2)r_h}\equiv\wmax~.
\label{wmax}
\end{equation}
Since $\tw r_0=r_0\sqrt{\omega^2-m_0^2}\simeq x_{\nu,k}$ from the mirror boundary condition, one has
\begin{equation}
 r_0>\frac{x_{\nu,k}}{\sqrt{\wmax^2-m_0^2}}\equiv \r0min~.
\label{r0min}
\end{equation}
For simplicity we only consider the case of $k=1$.
The size of the horizon radius $r_h$ is $r_h\simeq (2-6)\times 10^{-4}~{\rm fm}$ for $n=2,\cdots,7$ and $a_*=0.5,~1.0,~1.5$,
and $\r0min$ is roughly around $\sim 10\times r_h$ for typical values of other parameters.
In the present analysis we only consider the case of $j=m=1$ and $m_0=m_\pi=140$ MeV for simplicity.
Under this circumstance, we found that $\r0min/r_h=11.2,~8.99,~9.74$ and 
$\omega_{\rm max}r_h=0.400,~0.500,~0.462$ GeV for $a_*=0.5,~1.0,~1.5$.
\par
\begin{figure}
\includegraphics{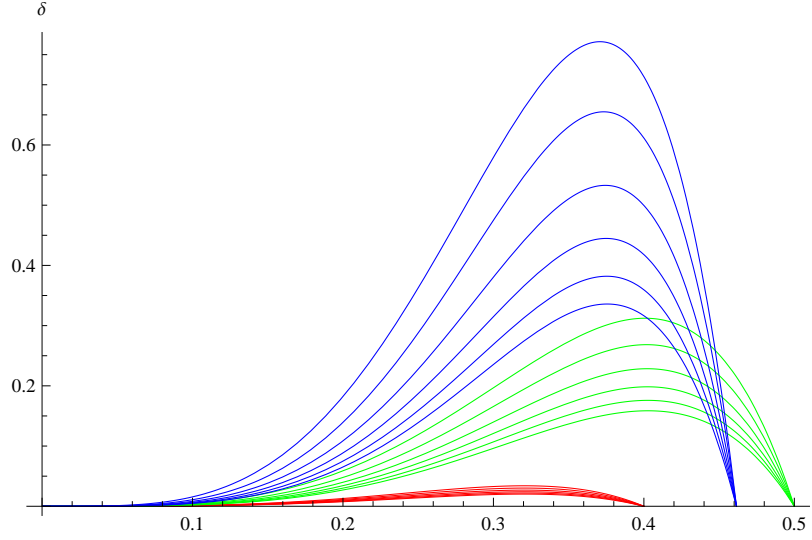}
\caption{\label{delta}Plots of $\delta$ in GeV as a function of $\omega r_h$ for $a_*=0.5$ (red), $a_*=1.0$ (green), $a_*=1.5$ (blue).
Each stacks of graphs corresponds to $n=2,\cdots,7$ from top to bottom.}
\end{figure}

Figure \ref{delta} shows $\delta$ as a function of $\omega r_h$ for different values of $a_*$.
In general, $\delta$ gets smaller as $n$ grows. Table \ref{time} shows some physical parameters for different values of $n$ and $a_*$.
Here $\omega_c$ is the value of $\omega$ which makes $\delta$ maximum, 
and $t_c=1/\delta_{\rm max}$ is the $e$-folding time.
\begin{table}
\begin{tabular}{cc|cccccc}\hline
&$n$ &$2$&$3$&$4$&$5$&$6$&$7$\\\hline
&$~a_*=0.5~$&$~2.85~$&$~3.73~$&$~4.52~$&$~5.23~$&$~5.87~$&$~6.48~$\\
$r_h~$&$~a_*=1.0~$&$2.43$&$3.32$&$4.11$&$4.83$&$5.49$&$6.11$\\
{\small (in $10^{-4}$ fm)}&$~a_*=1.5~$&$2.07$&$2.94$&$3.73$&$4.46$&$5.13$&$5.75$\\\hline
&$~a_*=0.5~$&$~0.0340~$&$~0.0302~$&$~0.0268~$&$~0.0241~$&$~0.0219~$&$~0.0202~$\\
$\delta_{\rm max}~$&$~a_*=1.0~$&$0.312$&$0.268$&$0.228$&$0.198$&$0.176$&$0.159$\\
{\small (in GeV)}&$~a_*=1.5~$&$0.772$&$0.656$&$0.533$&$0.445$&$0.382$&$0.336$\\\hline
&$~a_*=0.5~$&$19.3$&$21.8$&$24.6$&$27.3$&$30.0$&$32.6$\\
$t_c$ &$~a_*=1.0~$&$2.11$&$2.45$&$2.88$&$3.32$&$3.74$&$4.15$\\
(in $10^{-24}$ sec) &$~a_*=1.5~$&$~0.853~$&$~1.00~$&$~1.24~$&$~1.48~$&$~1.72~$&$~1.96~$\\\hline
&$~a_*=0.5~$&$0.321$&$0.321$&$0.321$&$0.321$&$0.321$&$0.321$\\
$\omega_c r_h$ &$~a_*=1.0~$&$0.401$&$0.402$&$0.402$&$0.402$&$0.403$&$0.403$\\
 &$~a_*=1.5~$&$~0.371~$&$~0.373~$&$~0.374~$&$~0.375~$&$~0.375~$&$~0.376~$\\\hline
&$~a_*=0.5~$&$14.0$&$14.0$&$14.0$&$14.0$&$14.0$&$14.0$\\
$r_0(\omega_c)/r_h$ &$~a_*=1.0~$&$11.3$&$11.3$&$11.3$&$11.3$&$11.3$&$11.3$\\
 &$~a_*=1.5~$&$~12.4~$&$~12.3~$&$~12.3~$&$~12.3~$&$~12.3~$&$~12.2~$\\\hline
\end{tabular}
\caption{\label{time}Some parameters for different values of $n$ and $a_*$.}
\end{table}
\par
For every $t_c$, the field amplitude is amplified by $e$-times, and the energy gets greater by an amount of 
$e^2=7.4$.
But the lifetime of a mini black hole $\tBH$ at the LHC is typically very short \cite{Argyres,Kanti}
\begin{equation}
 \tBH\sim\frac{1}{M_D}\left(\frac{\MBH}{M_D}\right)^{\frac{n+3}{n+1}}=(1.7-0.5)\times 10^{-26}~{\rm sec}~,
\label{tBH}
\end{equation}
for $n=1-7$, while $t_c\sim 10^{-23}-10^{-24}$ sec as shown in Table \ref{time}.
If the mini black hole lifetime is really short as Eq.\ (\ref{tBH}), the amplification by the superradiance is very small,
 $\lesssim\exp(1/100)$.
However, there is a possibility that $\tBH$ might be much longer than Eq.\ (\ref{tBH}), $\tBH\sim 10^{-17}$ sec by
\cite{Casadio}.
If the black hole lifetime is, for example, as long as $\tBH\sim 10^{-23}$ sec, then the amplitude can be amplified by 
$\sim e-e^{10}(\simeq 22000)$ times.
\par
To be more specific, consider the first law of thermodynamics of black holes, $\Delta\MBH=\Omega\Delta J$ where
$\Omega=a/(r_h^2+a^2)$ is the black hole angular velocity \cite{Cardoso}.
From Eq.\ (\ref{J}),
\begin{equation}
 \frac{\Delta\MBH}{\MBH}=\left(\frac{2}{n+2}\right)\frac{a_*}{1+a_*^2}\Delta a_*~.
\end{equation}
One can see from Eqs.\ (\ref{wmax}) and (\ref{r0min}) that $\r0min$ is minimum at $a_*=1$.
As the scattering process proceeds, the black hole loses its rotational energy and $a_*$ gets smaller.
If $a_*\le 1.0$ at the beginning, then $\r0min$ gets larger as $a_*$ becomes smaller.
As an example, let $a_*=0.5$ and $r_0=12 r_h$ at the beginning.
Note that $\r0min(a_*=0.5)=11.2 r_h$ as mentioned before.
During the scattering $a_*$ decreases while $\r0min$ increases, and eventually there is a point where $\r0min(a_*)=12 r_h$.
Since the scattering has started from $r_0=12 r_h$, the superradiance stops when $\r0min>12 r_h$.
This happens when $a_*=0.45$ where $\r0min(a_*=0.45)=12 r_h$.
Thus in this case the energy fraction transferred from the black hole is $\Delta\MBH/\MBH=1\%$ for $n=2$.
\par
If the scalar mass is heavier, then $\delta_{\rm max}$ gets smaller.
Qualitatively this is plausible because heavier particle would be more difficult to be amplified.
More thorough examinations will be given elsewhere \cite{jplee}.
\section{Conclusions}
In this work we analyzed the superradiant scattering by rotating black holes with a mirror ({\em black-hole bomb}) 
in higher dimensions which can be possibly produced at the LHC.
We found that for the pion the $e$-folding time is typically $10^{-23}-10^{-24}$ sec and $\r0min\sim 10 r_h$.
If the lifetime of the mini black hole is rather longer than the usually expected one, one can expect a huge amplification
of the energy of the scattered particle.
Present analysis is devoted to the case where the particle emission occurs only on the brane, but the generalization to
the bulk emission is quite straightforward \cite{jplee}.
\begin{acknowledgments}
This work was supported by the Basic Science Research Program through the National Research Foundation of Korea (NRF) 
funded by the Korean Ministry of Education, Science and Technology (2009-0088396).
\end{acknowledgments}

\end{document}